\documentclass[final]{svjour3}
\usepackage{graphicx}
\usepackage{rotating}
\usepackage{amssymb}
\usepackage{mathptmx}
\usepackage{float}
\usepackage[numbers]{natbib}
\usepackage{array}
\usepackage{makecell}
\makeatletter
\journalname{Journal of Low Temperature Physics}

\bibpunct{}{}{,}{s}{}{,}

\begin{document}
\newcommand{\hdblarrow}{H\makebox[0.9ex][l]{$\downdownarrows$}-}
\title{The CROSS Experiment: Rejecting Surface Events by PSD Induced by Superconducting Films}

\author{H.~Khalife$^1$ \and L.~Bergé$^1$ \and M.~Chapellier$^1$ \and L.~Dumoulin$^1$ \and A.~Giuliani$^{1,2}$ \and P.~Loaiza$^3$ \and P.~de Marcillac$^1$ \and S.~Marnieros$^1$  \and C.A.~Marrache-Kikuchi$^1$ \and C.~Nones$^4$ \and V.~Novati$^1$ \and E.~Olivieri$^1$ \and Ch.~Oriol$^1$ \and D.V.~Poda$^1$ \and Th.~Redon$^1$ \and A.S.~Zolotarova$^1$ }

\institute{$^{1}$ CSNSM, Univ. Paris-Sud, CNRS/IN2P3, Université Paris-Saclay, 91405 Orsay, France\\ 
$^{2}$ DISAT, Università dell’Insubria, I-22100 Como, Italy\\
$^{3}$ LAL, Univ. Paris-Sud, CNRS/IN2P3, Université Paris-Saclay, 91405 Orsay, France\\
$^{4}$ IRFU, CEA, Université Paris-Saclay, F-91191 Gif-sur-Yvette, France\\
\\
\email{hawraa.khalife@csnsm.in2p3.fr}}

\maketitle

\begin{abstract}

Neutrinoless double beta ($0\nu\beta\beta$) decay  is a hypothetical rare nuclear transition ($T_{1/2}>10^{25}-10^{26}$ y). Its observation would provide an important insight about the nature of neutrinos (Dirac or Majorana particle) demonstrating that the lepton number is not conserved. This decay can be investigated with bolometers embedding the double beta decay isotope ($^{76}$Ge, $^{82}$Se, $^{100}$Mo, $^{116}$Cd, $^{130}$Te...), which perform as low temperature calorimeters (few tens of mK) detecting particle interactions via a small temperature rise read out by a dedicated thermometer. CROSS (Cryogenic Rare-event Observatory with Surface Sensitivity) aims at the development of bolometric detectors (based on Li$_{2}$MoO$_{4}$ and TeO$_{2}$ crystals) capable of discriminating surface $\alpha$ and $\beta$ interactions by exploiting superconducting properties of Al film deposited on the detector surface. We report in this paper the results of tests on prototypes performed at CSNSM (Orsay, France) that showed the capability of a-few-$\mu$m-thick superconducting Al film deposited on crystal surface to discriminate surface $\alpha$ from bulk events, thus providing the detector with the required PSD (pulse shape discrimination) capability. The CROSS technology would further improve the background suppression and simplify the detector construction (no auxiliary light detector is needed to reject alpha surface events) with a view to future competitive double beta decay searches.


\end{abstract}

\section{Introduction}

Neutrinoless double beta decay consists in the transformation of an even-even nucleus ($A$,~$Z$) into a  lighter isobar ($A$,~$Z+2$) accompanied by the emission of 2 electrons and no other particle [1]. This process is forbidden within the Standard Model since its discovery would confirm that lepton number is not a symmetry of nature. In addition, it will imply that neutrinos are the only fermions that coincide with their antiparticle, i.e. Majorana particles. This process is energetically allowed for 35 nuclei ($^{76}$Ge, $^{82}$Se, $^{100}$Mo, $^{116}$Cd, $^{130}$Te, $^{136}$Xe ...) [2] and its signature is a monochromatic peak at the $Q$-value ($Q_{\beta\beta}$) of the transition corresponding to the sum of the kinetic energies carried by the two emitted electrons. The most promising and desirable isotopes to study are those with a high $Q_{\beta\beta}$ above the natural $\gamma$ radioactivity endline ($^{208}$Tl, ~ 2.615 MeV), since this will imply less background in the region of interest (ROI) and higher decay probability (since the phase-space of the process increases with higher $Q_{\beta\beta}$). Another important consideration is the natural abundance for the $0\nu\beta\beta$ isotope. High isotopic abundance (I.A.) will translate into easiness of enrichment of the $\beta\beta$ isotope and a lower cost as well. The current most stringent half-life limits on $0\nu\beta\beta$ decay are of the order of 10$^{25}$–10$^{26}$ y [3]. This makes any experiment aiming at searching for this rare decay extremely challenging.
An experiment has to be performed with radiopure large-mass high-resolution detectors and negligible radioactivity from anything except the nucleus under study, because otherwise the signal will be buried in the background.

\section{Bolometric technique and choice}
Bolometers are very powerful nuclear detectors to study $0\nu\beta\beta$ [4]. When a particle releases energy in the absorber (crystal) it produces lattice vibrations that generate a temperature rise in the crystal measured by a thermometer glued on the crystal surface. Working at cryogenic temperature is crucial to lower the heat capacity of the insulating crytal ({\it C~$\propto$~T$^3$}), because otherwise the signal amplitude will be dimmed. Bolometers offer wide choices of absorber materials making it possible to improve the sensitivity of the experiment through the maximization of the efficiency and the minimization of the background by choosing a compound compatible with an active background-rejection technique (pulse shape discrimination or detection of scintillation [5] or Cherenkov light [6]). The CROSS choice of the isotopes and the compound crystals is based on years of dedicated R$\&$Ds. Li$_{2}$MoO$_{4}$ [7,8] and TeO$_{2}$ [9] were found to be among the most promising detector materials, containing the $0\nu\beta\beta$ isotopes $^{100}$Mo ($Q_{\beta\beta}$~=~3034 keV $\&$ I.A. = 9.6$\%$) and $^{130}$Te ($Q_{\beta\beta}$ = 2527 keV $\&$ I.A. = 34$\%$) respectively. $^{100}$Mo fulfills the major requirement of a high $Q_{\beta\beta}$ above 2.615 MeV. $^{130}$Te is an attractive candidate for being the $0\nu\beta\beta$ isotope with the highest natural abundance. Li$_{2}$MoO$_{4}$ and TeO$_{2}$ have successfully been studied in LUMINEU and CUORE projects respectively and shown to have a high energy resolution in ROI ($\bigtriangleup_{FWHM}$ = 5 keV) and high internal radiopurity as an effective purification proccesses can be developed for these compounds [6,8,10].

\section{Background rejection technique}
A source of background, as mentioned in  the Introduction, comes from the highest intensity natural $\gamma$ line from $^{208}$Tl (2.615 MeV) in the $^{232}$Th decay chain, which can provide an important $\gamma$ background contribution in the $^{130}$Te ROI. Above 2.615 MeV, the $\gamma$ background contribution can come only from weak lines of $^{214}$Bi and it can be made negligible. Energy-degraded $\alpha$s coming from $^{238}$U and $^{232}$Th radioactive chains due to surface contamination of the crystal and the surrounding materials (copper holder, Polytetrafluoroethylene...) contribute to the background in  ROI in both crystals [8]. In addition, there is a contribution from $\beta$s emitted near the surface from the surrounding materials. Surface radioactive contamination is considered the most challenging limitation to the sensitivity for $0\nu\beta\beta$ bolometeric experiments because of the non-existence of a dead layer for bolometers. The rejection of $\alpha$s can be achieved by exploiting the scintillation and Cherenkov radiation (for Li$_{2}$MoO$_{4}$ and TeO$_{2}$ respectively) emitted by the absorber and detected by a separate light detector (which is also a bolometer made out of Ge wafers) facing the crystal, since $\alpha$s and $\beta$s have different light yield.\\

CROSS proposes another technique of surface background rejection which does not require an additional device. The key idea is to provide the crystal surface with pulse shape modification capability (in terms of the pulse rise-time for example), so that it would be possible to discriminate surface $\alpha$s from the bulk events ($\gamma/\beta$), which include the $0\nu\beta\beta$ signal. This is obtained by depositing a thin layer (few $\mu$m) of superconducting Al film (critical temperature $T_{c}$ = 1.2 K) via evaporation. The superconducting Al film acts as a pulse shape modifier for surface events up to $\sim$ 1 mm depth [18] from the crystal surface coated by the film. The feasibility of this method has been proven in a test done back in 2010 (CSNSM, Orsay) with a TeO$_{2}$ having NbSi thin film (Anderson insulator) [11,12] as a fast thermal sensor ( $\sim$ 1 ms time response).

The tests reported here on small crystals of Li$_{2}$MoO$_{4}$ and TeO$_{2}$ ($20\times20\times10$ mm$^3$) with one surface covered with a few-$\mu$m-thin Al film ($18\times18$ mm$^2$) show that the discrimination capability between surface and bulk events is successfully achieved when using NTD (neutron transmutation doped) Ge thermistor as a sensor (slow sensor) [13]. The NbSi and the NTD sensors showed to work in a different way. The NbSi film is deposited ($13\times13$ mm$^2$) directly on the crystal over a large surface ($14\times14$ mm$^2$, Fig. 1 top right). This makes NbSi sensor sensitive to the prompt athermal component of the phonon population produced by the impinging particle, while NTDs are sensitive rather to the thermal component due to their intrinsic slowness and the glue interface. When a particle releases energy in the absorber close to the Al-film-coated surface it generates high energy phonons that break Cooper pairs in the superconducting film. This effect is less important for bulk events, since the generated phonons reach the surface with a lower average energy due to the quasi-diffusive mode of phonon propagation [14]. The bulk events will be felt directly by the NbSi sensor and later after thermalization by the NTD sensor. For the surface events, a significant fraction of the particle energy will be trapped in the superconductive film in the form of quasiparticles for a few millisecond due to the long recombination time of quasiparticles at a temperature well below the transition energy of the superconducting Al film. The subsequent recombination of quasiparticles to a lower energy $\sim$ 1.2 K phonons will add a delayed component to the signal when read by the NbSi (making bulk events faster than surface events), while it will accelerate the detection in case of the thermal phonon sensitive sensor (NTD) with respect to the bulk events due to the faster thermalization (making surface events faster than bulk events).

\begin{figure}[htbp]
\centering
\includegraphics[width=0.44\linewidth]{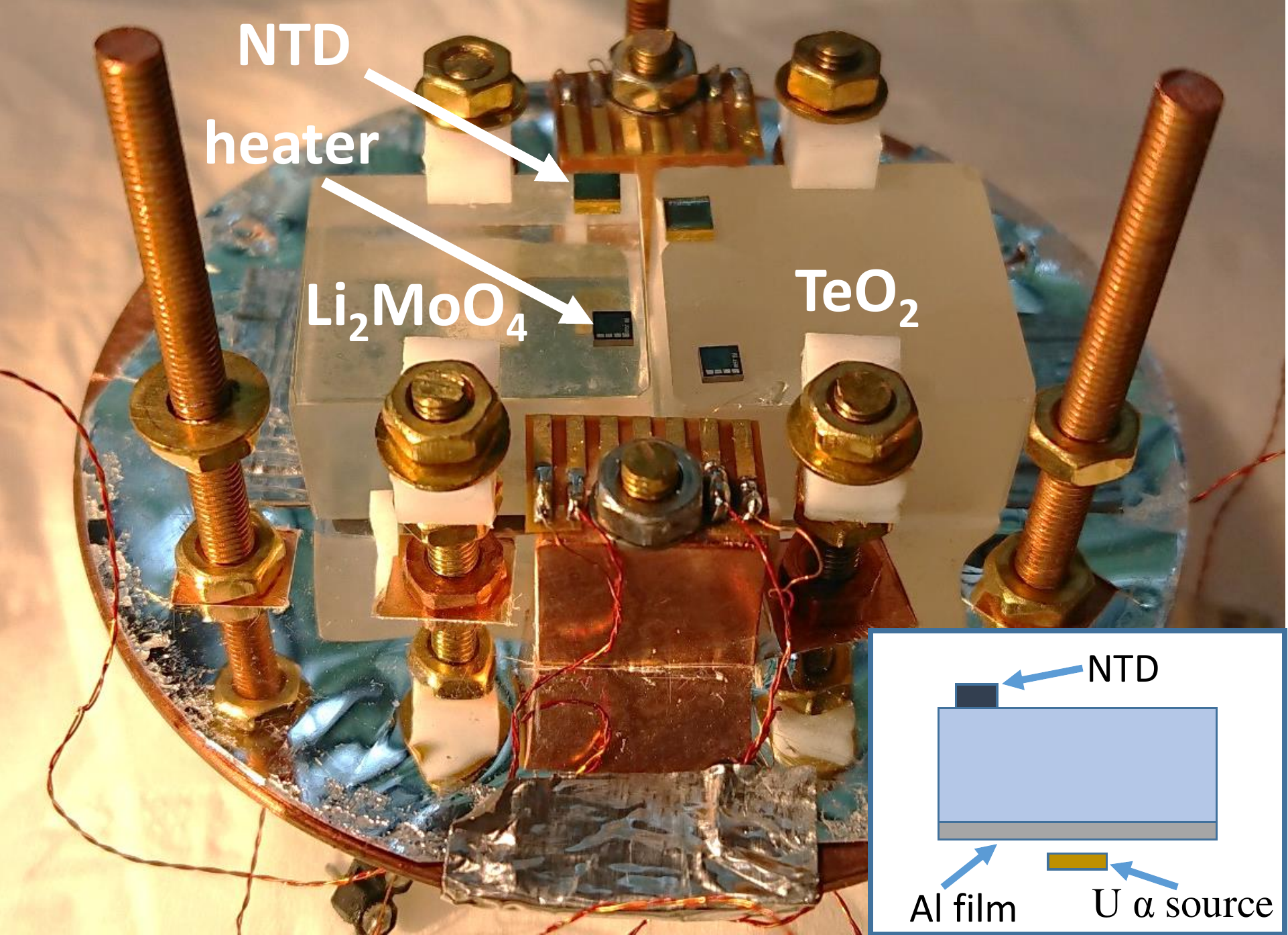}
\includegraphics[width=0.365\linewidth]{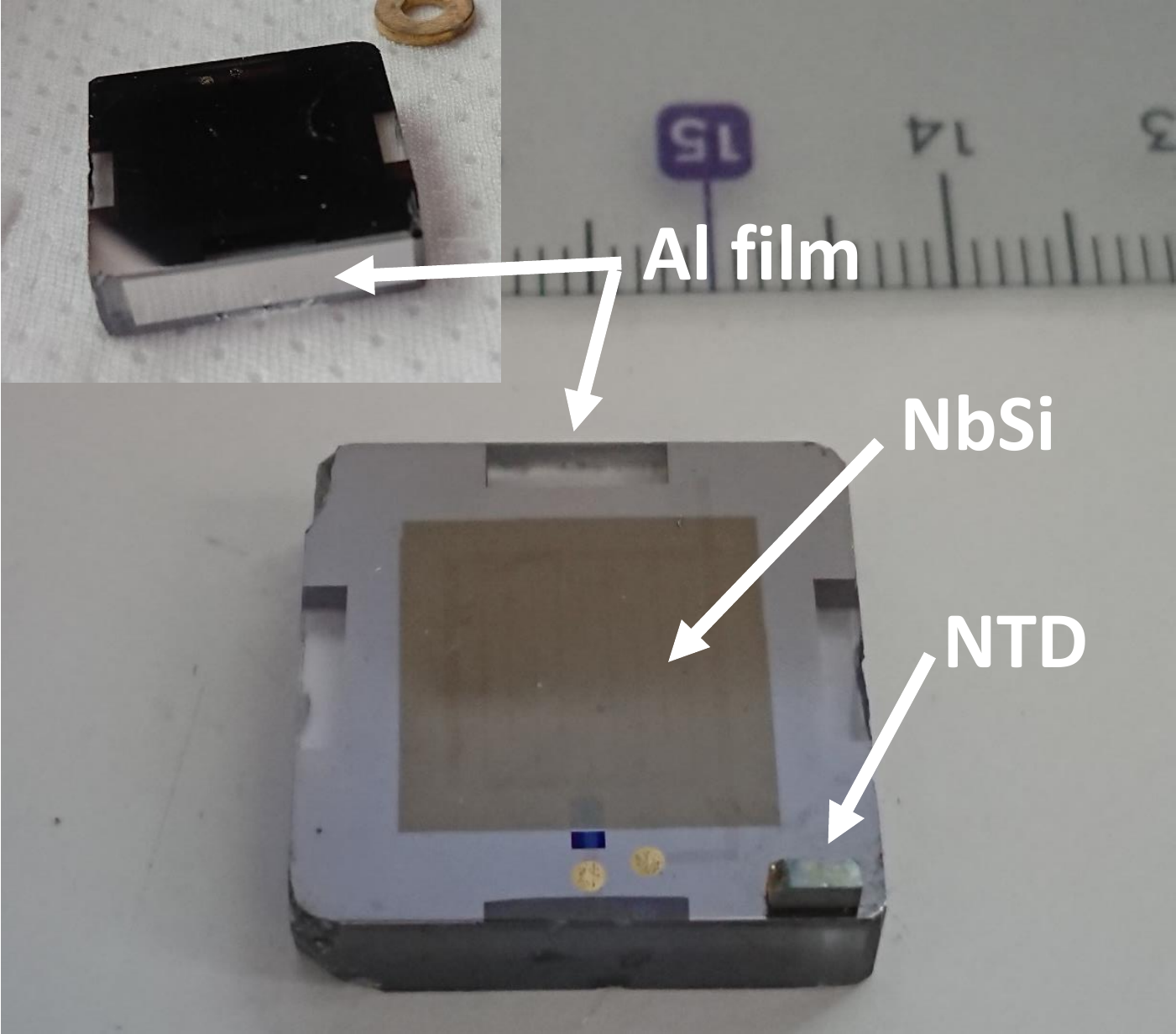}
\\
\includegraphics[width=0.3\linewidth]{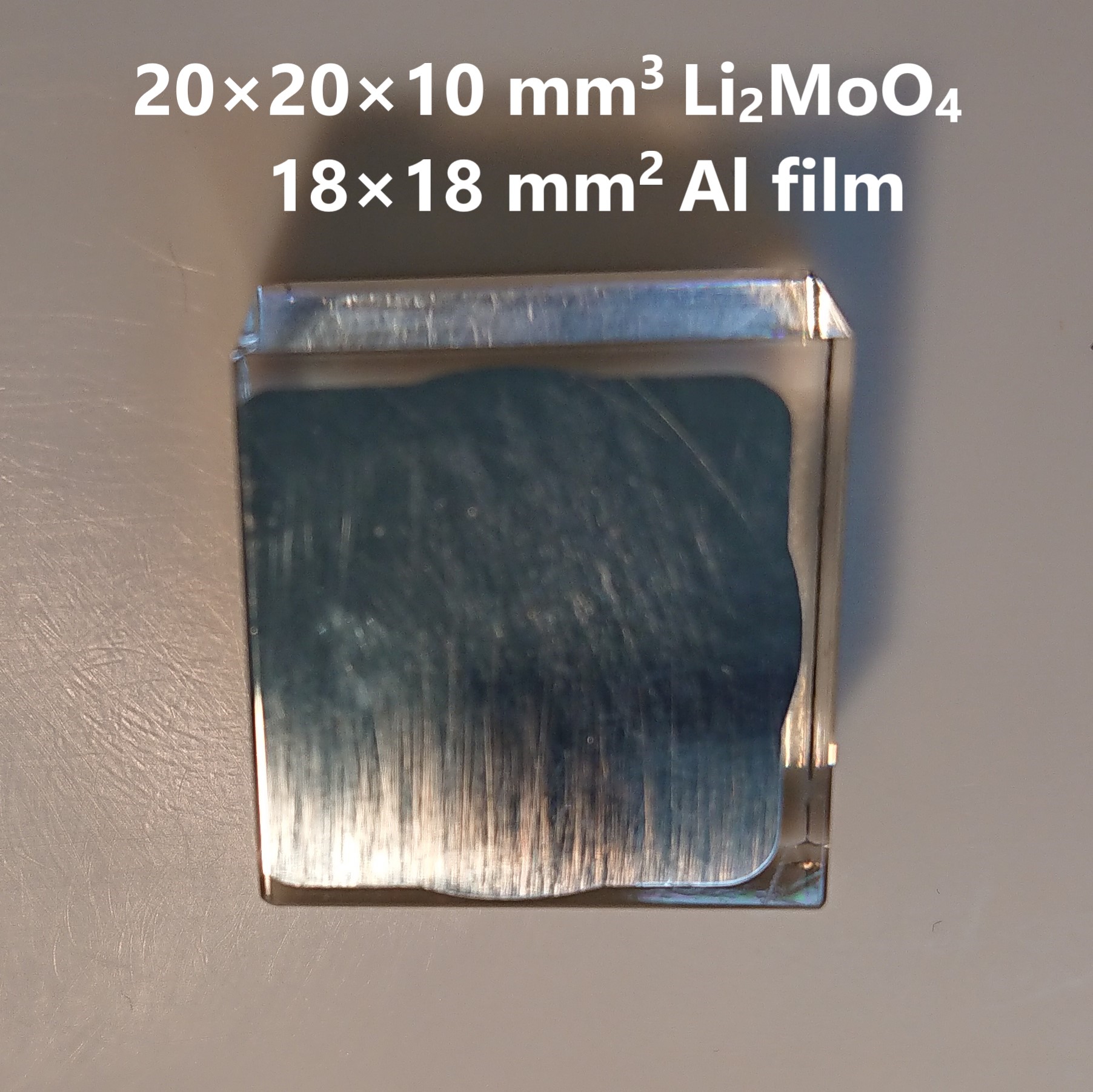}
\includegraphics[width=0.285\linewidth]{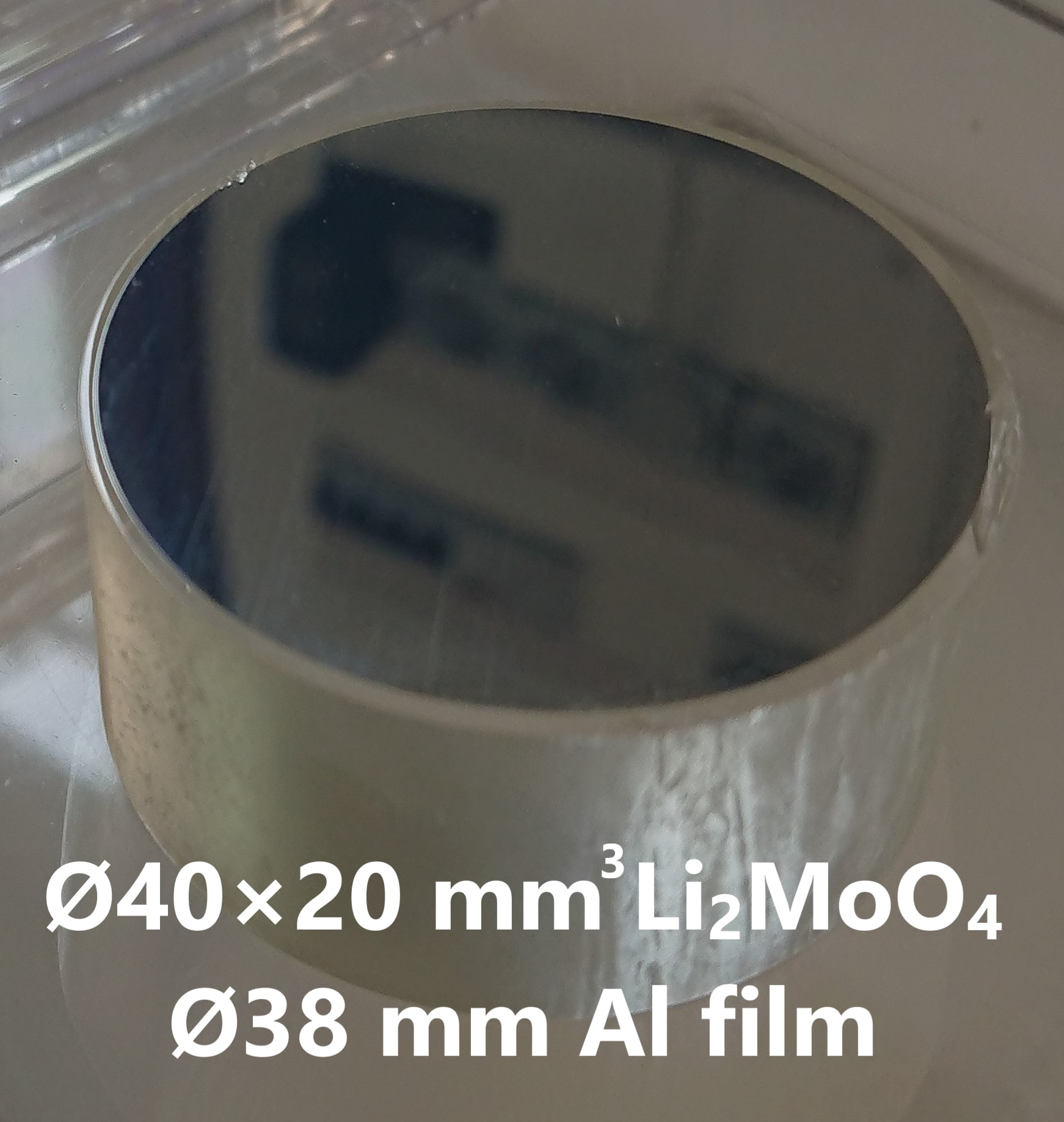}
\includegraphics[width=0.3\linewidth]{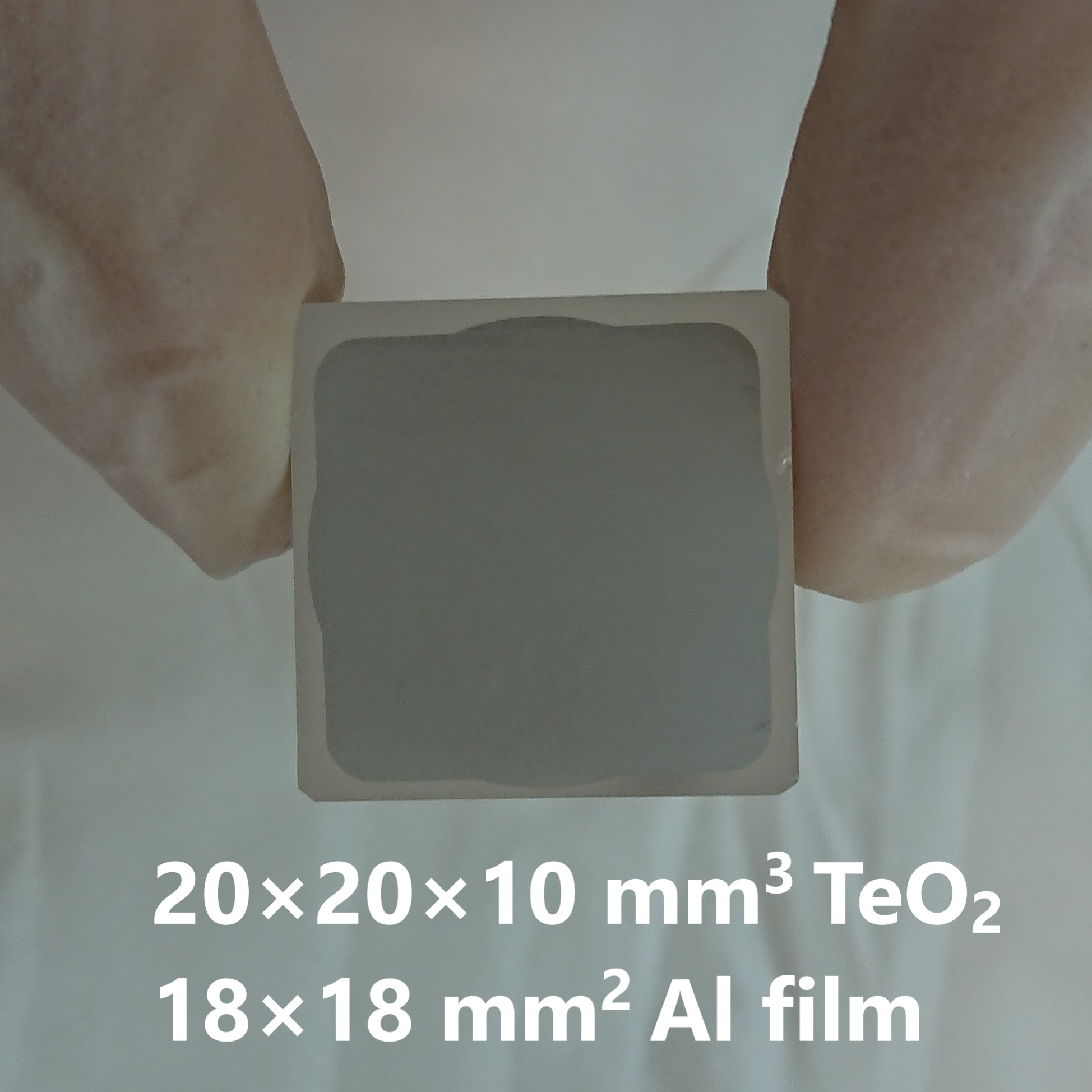}
\linebreak
\caption{({\it top left}) Photograph of the typical assembly of CROSS detectors, Li$_{2}$MoO$_{4}$ and TeO$_{2}$ (bolo1 and bolo3, see Table 1). ({\it top right}) The  TeO$_{2}$ crystal having two readouts NbSi and NTD Ge (bolo5), and 10 $\mu$m Al film on one 20$\times$5 mm$^{2}$ side. ({\it bottom left}) Small Li$_{2}$MoO$_{4}$ crystal with 10 $\mu$m Al film evaporated on 20$\times$20 mm$^{2}$ side. ({\it bottom middle})  Large Li$_{2}$MoO$_{4}$ crystal (bolo2) with the 10 $\mu$m Al film evaporated on one circular side (on the other side we glue an NTD and a heater). ({\it bottom right}) TeO$_{2}$ crystal with 10 $\mu$m Al film evaporated on 20$\times$20 mm$^{2}$ side. All the faces with Al film are facing a uranium $\alpha$ source (surface events) and for bolo5 another uranium $\alpha$ source deposited directly on the crystal surface with no Al film (the other  20$\times$5 mm$^{2}$ side without Al film).}
\label{fig:one}
\end{figure}

\section{Experimental results}
\subsection{Assembly description}
We have performed several runs on different bolometers (Fig.~\ref{fig:one} and Table~\ref{table:bolo}) with different sensors, film types and thicknesses to achieve the best discrimination capability. A thin film of Al was evaporated on a surface of the bolometers using a dedicated evaporator. An NTD Ge thermistor was used for all the  bolometers. It was glued on a crystal surface with either six 25-$\mu$m-thick glue spots or a thin layer of glue spread all under the NTD (bolo2). In addition, a heater was glued to provide periodically thermal pulses in order to stabilize the response of the bolometers [15]. The crystals were fixed on a copper plate by means of PTFE pieces and brass/copper screws. A uranium $\alpha$ source was placed facing the surface with Al film to generate surface events. The source provides two main $\alpha$ lines at $\sim $ 4.2~MeV and $\sim $~4.7~MeV, and also $\beta$ particles from the isotope $^{234m}$Pa, with a spectrum extending up to $\sim $ 2.27 MeV. The TeO$_{2}$ with NbSi (bolo5) sensor tested in the past [16] was tested again with a modification. An NTD was added as a second sensor to appreciate the different response between two sensors on the same crystal with sensitivities depending on the phonon energy. This crystal has $10~\mu$m Al film faced by a uranium $\alpha$ source (surface events) and another uranium $\alpha$ source deposited directly on the opposite surface to emulate bulk events (there is no pulse shape difference between $\alpha$ and $\beta/\gamma$ for  TeO$_{2}$, while such difference is expected for Li$_2$MoO$_4$ as often happens in bolometers based on crystal scintillators [17,4]).

\begin{table}[t]
\centering 
\begin{tabular}{c c c c c c c c} 
\hline 
 & Bolometer & Sensor (glue) & \thead{Crystal size \\ (mm$^{3}$)} & \thead{Crystal \\ mass (g)} &  \thead{Al film \\ thickness} & DP & \thead{Sensitivity \\ (nV/keV)} \\ 
[0.5ex] 
\hline 

bolo1 & Li$_{2}$MoO$_{4}$ & NTD Ge (6 spots)   & 20$\times$20$\times$10    & 12   &  10 $\mu$m  &  12 & 53 \\

bolo2 & Li$_{2}$MoO$_{4}$  & NTD Ge (veil)  & $\varnothing$40$\times$20   & 67 & 10 $\mu$m & 7 & 37 \\

bolo3 & TeO$_{2}$   & NTD Ge (6 spots)   & 20$\times$20$\times$10   & 25 & 1 $\mu$m  & 3 & 43\\

bolo4 & TeO$_{2}$   & NTD Ge (6 spots)   & 20$\times$20$\times$10   & 25 &  10 $\mu$m & 4 & 44 \\

bolo5 &TeO$_{2}$    & \thead{NTD Ge (2 spots)  \\+\\ NbSi (evaporation)}  & 20$\times$20$\times$5  & 12.5 & 10 $\mu$m & & \thead{70 \\ \\ 54}   \\

\hline 
\end{tabular}
\linebreak
\caption{Description of the five bolometers that have been tested and their respective DP (Eq. (1)) and sensitivity at 22~mK (bolo5 was operating at 30 mK). A uranium $\alpha$ source irradiated the film in all the detectors. DPs for bolo 5 are missing because the alpha population has a large spread presumably due to position effects. As a results, the distribution of the rise-time is strongly not gaussian and the DP as defined here becomes meaningless.}
\label{table:bolo} 
\end{table}

\begin{figure}[t]
\centering
\includegraphics[width=0.47\linewidth]{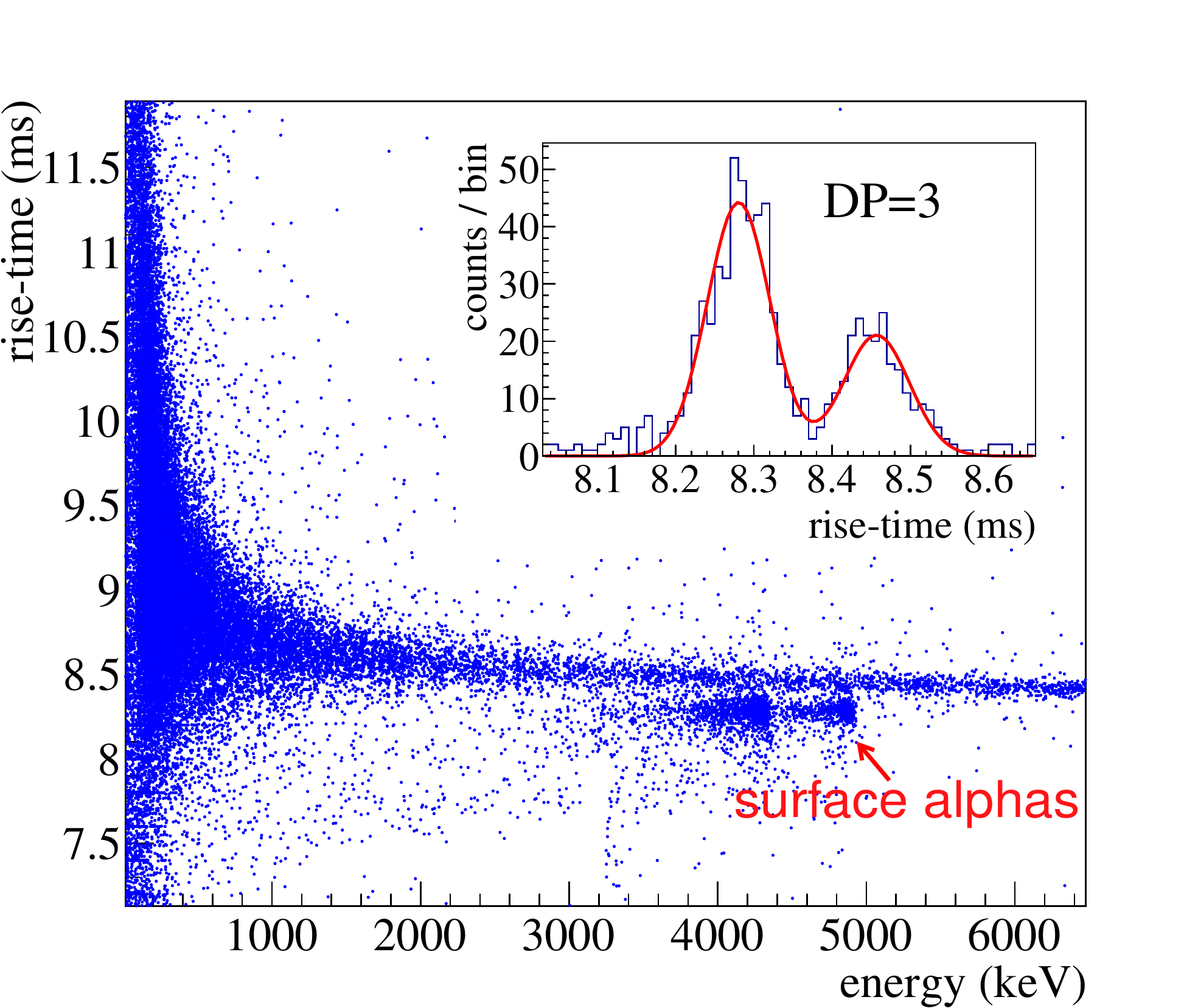}
\includegraphics[width=0.43\linewidth]{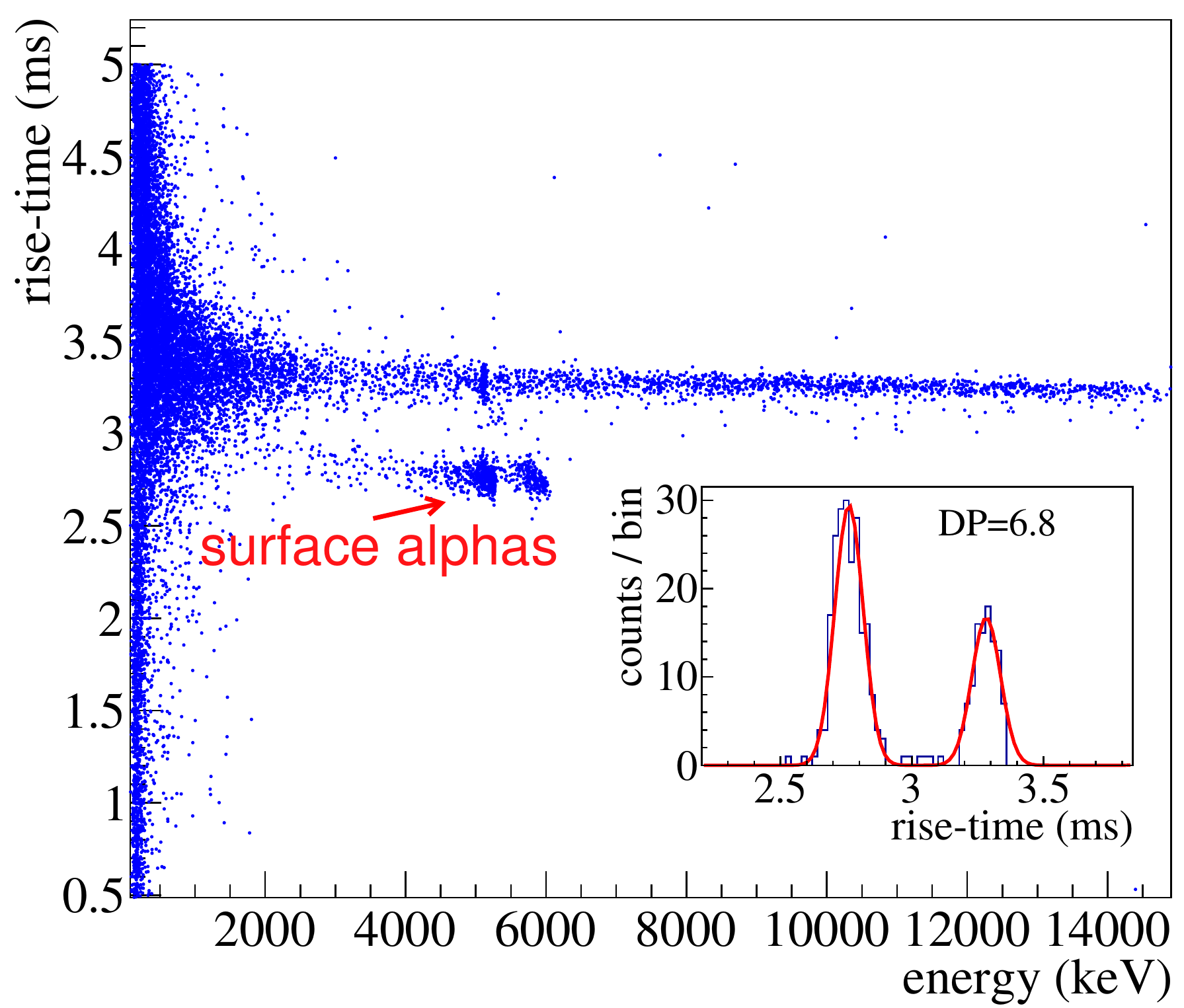}
\linebreak
\caption{ The rise-time is plotted as a function of the energy: ({\it left}) TeO$_{2}$ (bolo3) with 1 $\mu$m Al film;  ({\it right}) Li$_{2}$MoO$_{4}$ (bolo2) with 10 $\mu$m Al film.}
\label{fig:two}
\end{figure}

\begin{figure}[t]
\centering
\includegraphics[width=0.45\linewidth]{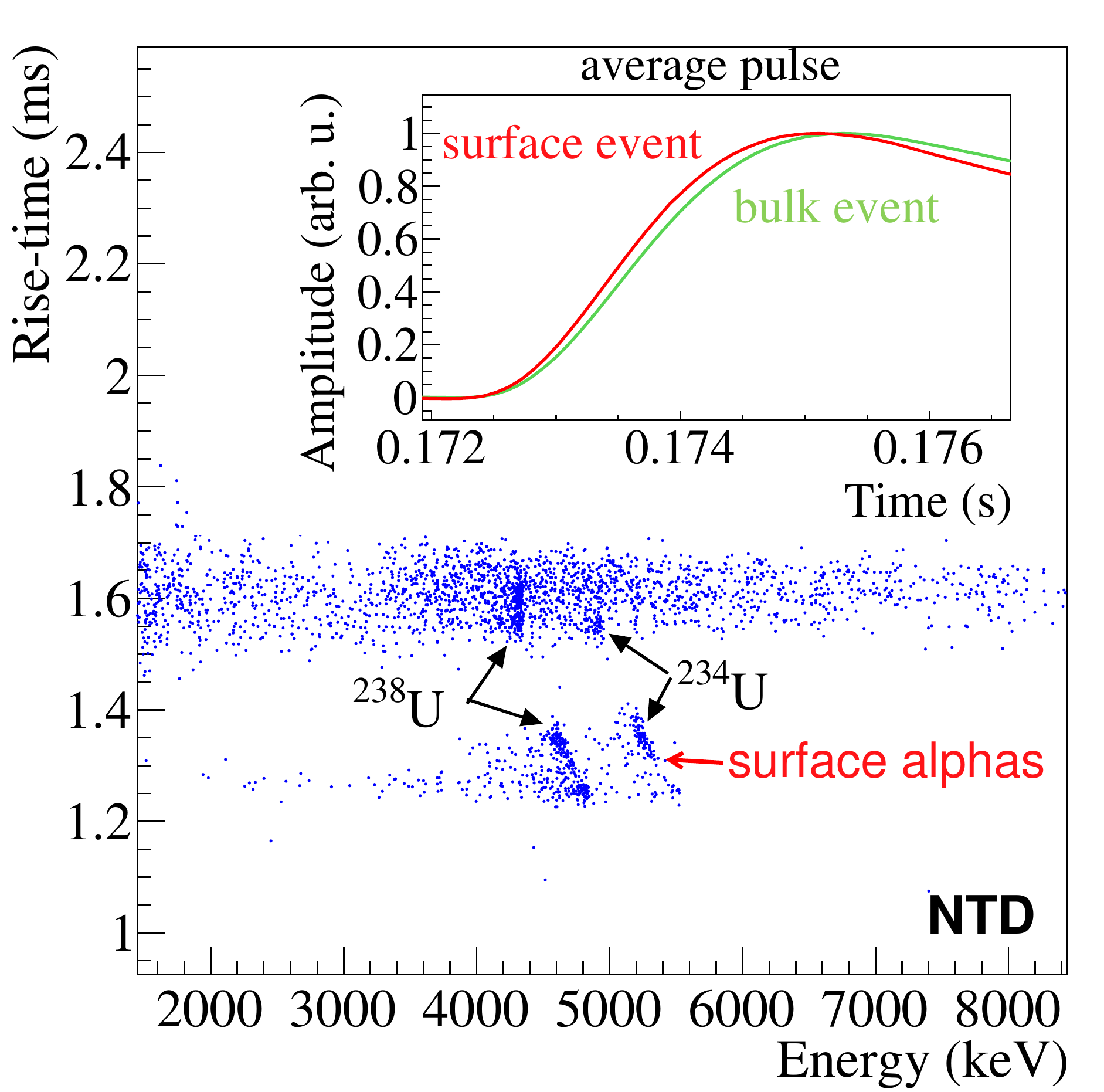}
\includegraphics[width=0.446\linewidth]{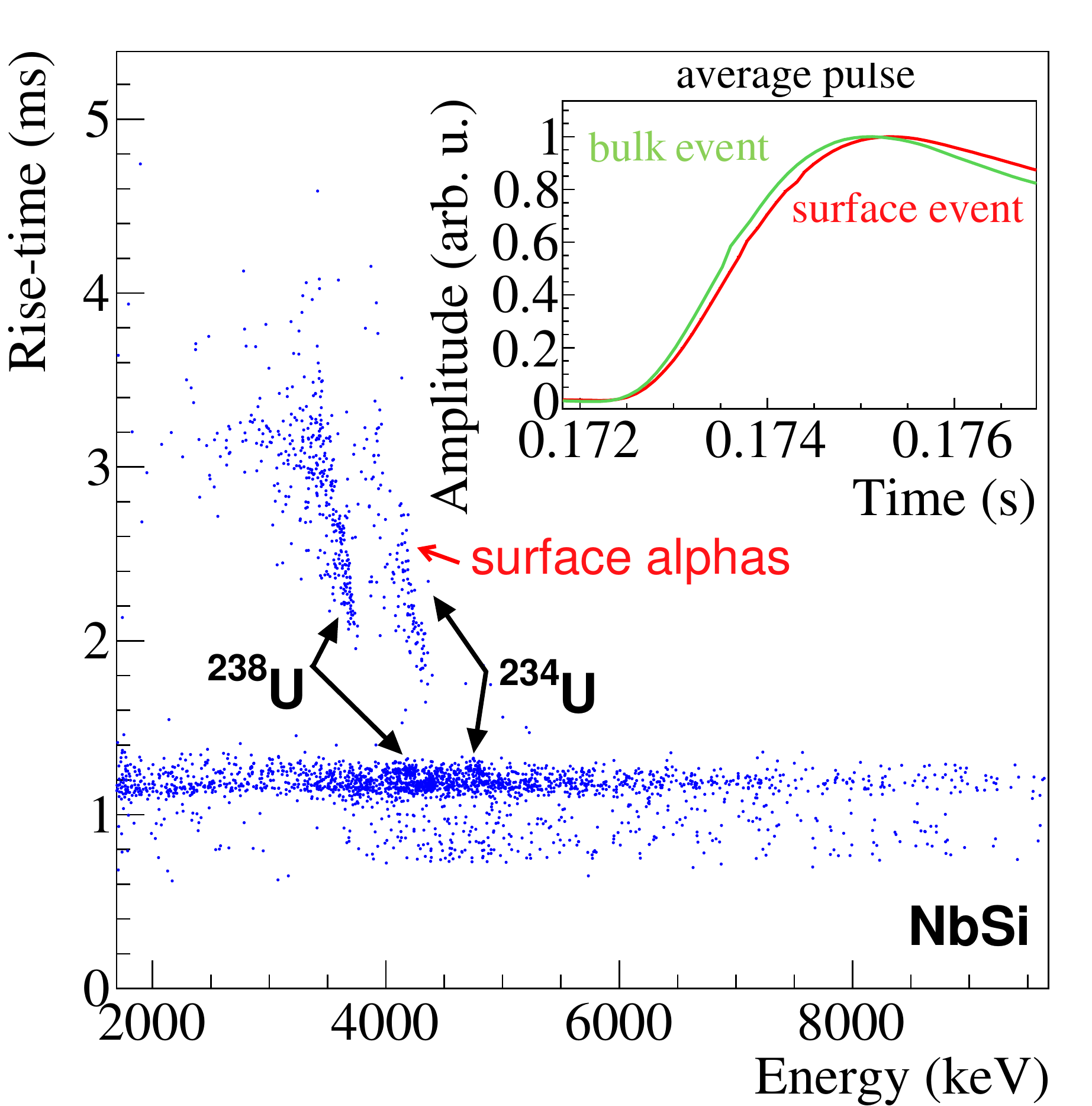}
\linebreak
\caption{ The rise-time is plotted as a function of the energy: TeO$_{2}$ with two readouts, NTD Ge ({\it left}) and NbSi ({\it right}).}
\label{fig:three}
\end{figure}

\subsection{Pulse shape discrimination results}
The tests have shown that the surface $\alpha$ events can be safely rejected by their different pulse shape from the bulk events in all of the bolometer listed in Table ~\ref{table:bolo}, while the identification of surface $\beta$ particles requires additional R$\&$D work. The PSD parameter that we discuss here is the rise-time (time from 10$\%$ to 90$\%$ of the pulse maximum amplitude). Fig.~\ref{fig:two}  shows the rise-time parameter plotted as a function of the deposited energy in the crystal for TeO$_{2}$ (bolo3) and Li$_{2}$MoO$_{4}$ (bolo2). The surface $\alpha$s (from the uranium source) are clearly separated from the bulk making their rejection possible. The discrimination capability can be quantified in terms of the discrimination power DP (Eq. (1)), where $\mu$ and $\sigma$ are the mean and standard deviation respectively extracted from the gaussian fit applied to the surface and bulk event distributions (insets of Fig.~\ref{fig:two}): 

\begin{equation}
	DP=\frac{ |{\mu_{surface}-\mu_{bulk}}|}{\sqrt{\sigma_{surface}^2+\sigma_{bulk}^2}} ,
\end{equation}

The DP values for all the bolometer are given in Table ~\ref{table:bolo}. Comparing the results of bolo3 and bolo4, which are identical but the film thickness, there is a weak but not conclusive indication that thicker films may lead to a better discrimination. In addition, results showed that depositing Al film does not affect the energy resolution and sensitivity of the bolometer [18].

Fig.~\ref{fig:three} shows the discrimination results for TeO$_{2}$ bolometer instrumented with NTD and NbSi. It is confirmed that surface events are faster (slower) than the bulk read out by the NTD Ge sensor (NbSi sensor) as discussed in section 3. The two alpha lines in the bulk (from the uranium source deposited on the crytsal surface directly) are at their nominal energies (4.2 MeV and 4.7 MeV), whereas the surface alphas are either energy-overestimated (NTD sensor) or energy-underestimated (NbSi sensor). The former observation is because of the better thermal conversion of Al film (a superconductor compared to the crystal which is an insulator), while the later effect is because of lower pulse amplitude caused by a trapped energy in form of quasiparticles. Note that the calibration in both cases was done with the low energy gamma quanta of 352 keV from $^{238}$U decay chain. One can notice that the plots of Fig.~\ref{fig:three} show a linear behaviour (PSD parameters don't depend on deposited energy) for bolo5 unlike bolo2 and bolo3 (Fig.~\ref{fig:two}) and this is because bolo5 was operating at a higher temperature. It has to be noticed that the U source, due to the geometry of the set-up, can produce also a small population of bulk-like (the reason of the double poplution in the bulk) events when $\alpha$ particles hit directly the TeO$_2$ crystal face, which is not fully covered by the Al film [16].

The discrimination power depends on the pulse parameter used. Another pulse shape parameter was tested and was found to provide a higher DP's than the rise-time (a full description of this parameter and the related results can be found in [18]). 

\section{Conclusions}
We have described in this paper the technology behind CROSS (Cryogenic Rare-event Observatory with Surface Sensitivity), which showed its capability of rejecting surface $\alpha$ background when a superconducting Al film is deposited on a side of the crystal. This method will allow us to simplify substantially the bolometric structure by getting rid of a light detector. Future goals will be to fully coat the crystals and operate a medium-scale demonstrator with tens of elements in the Canfranc underground laboratory (Spain).

\begin{acknowledgements}
The project CROSS is funded by the European Research Council (ERC) under the European-Union Horizon 2020 program (H2020/2014-2020) with the ERC Advanced Grant no. 742345 (ERC-2016-ADG). The PhD fellowship of H. Khalife has been partially funded by the P2IO LabEx (ANR-10-LABX-0038) managed by the Agence Nationale de la Recherche (France) in the framework of the 2017 P2IO doctoral call.
\end{acknowledgements}

\end{document}